\documentclass[journal]{IEEEtran}

{}
{}
{}
\usepackage{graphicx,amsmath,amssymb,amsfonts,xcolor,soul,circuitikz,subfigure,placeins, multirow,stfloats, booktabs, enumerate, comment}%
\usepackage{float}
\floatstyle{ruled}
\newfloat{model}{thp}{lop}
\floatname{model}{\textsc{Model}}

\def\CaseA{\mbox{\small{\textsc{\textsf{Case A}}}}}
\def\CaseB{\mbox{\small{\textsc{\textsf{Case B}}}}}
\def\CaseC{\mbox{\small{\textsc{\textsf{Case C}}}}}

\DeclareMathOperator*{\minimize}{minimize}

\def\ModelOne{\mbox{\small{\textsc{\textsf{Model \ref{Mod: UC Col}}}}}}

\def\CaseA{\mbox{\small{\textsc{\textsf{Case A}}}}}
\def\CaseB{\mbox{\small{\textsc{\textsf{Case B}}}}}
\def\CaseC{\mbox{\small{\textsc{\textsf{Case C}}}}}
\usepackage{ulem}
\begin{document}

\title{Economic Impact of Wind Generation Penetration in the Colombian Electricity Market}

\author{\IEEEauthorblockN{Alvaro Gonzalez-Castellanos, David Pozo, Sergio Mart\'inez, Luis L\'opez, Ingrid Oliveros}

\thanks{This paper is a preprint of a paper submitted to "IET Generation, Transmission \& Distribution". If accepted, the copy of record will be available at the IET Digital Library.
This work was supported by the strategic energy area 2016-2018 Universidad del Norte and theSkoltech NGP Program (Skoltech-MIT joint project).
A. Gonzalez-Castellanos, and D. Pozo are with the Center for Energy Systems, Skolkovo Institute of Science and Technology (Skoltech), Moscow, Russia; Sergio Mart\'inez is with the Universidad Politécnica de Madrid, Madrid, Spain; Luis L\'opez and Ingrid Oliveros are with the Electrical Engineering Department, Universidad del Norte, Barranquilla, Colombia. A. Gonzalez-Castellanos is the corresponding author (e-mail: alvaro.gonzalez@skolkovotech.ru).}}

\maketitle
\thispagestyle{plain}
\pagestyle{plain}

\begin{abstract}
The creation of the Renewable Energy Law (Law 1715 of 2014) promotes the introduction of large-scale renewable energy generation in the Colombian electricity market. The new legislation aims to diversify the country's generation matrix, mainly composed of hydro and fuel-based generation, with a share of 66\% and 34\% respectively. Currently, three wind generation projects, with an aggregated capacity of 500 MW, have been commissioned in the North of the country. This study analyses the economic impact of the large-scale introduction of wind generation on both, the market spot price and conventional generation plants operation. For this purpose, the study builds a unit commitment model to mimic the current market legislation and the system's generation data.  We show that the introduction of wind energy into the Colombian electricity market would impact the generation share of large hydro and gas-fired power plants. The hydro generation has an important role in balancing the generation for fluctuations on the wind resource. Meanwhile, the gas-fired plants would decrease their participation in the market, proportionally to the introduction of wind generation in the system, by as low as 20\% of its current operation. 
\end{abstract}

\section*{Nomenclature}\label{Sec: Nomenclature}
\addcontentsline{toc}{section}{Nomenclature}
\begin{IEEEdescription}[\IEEEusemathlabelsep\IEEEsetlabelwidth{$RD_{g},RU_{g}$}]
\item[\scalebox{1.2}{\textit{Abbreviations}}]
\item
\item[ASIC] Manager of the system for commercial exchanges
\item[COP] Colombian peso
\item[CREG] Energy and gas regulation commission
\item[DR] Demand response
\item[FNCER] Non-conventional renewable energy sources
\item[ITCZ] Inter-tropical convergence zone
\item[ISA] Transmission system operator
\item[OEF] Firm energy obligation
\item[PROURE] Program for the rational and efficient use of conventional and non-conventional energy
\item[REL] Renewable energy law (1715 of 2014)
\item[SIN] National interconnected system
\item[XM] Market Operator.
\item
\item[\scalebox{1.2}{\textit{Indices}}]
\item
\item[g] generation unit
\item[t] time period [h].
\item
\item[\scalebox{1.2}{\textit{Parameters}}]
\item
\item[$c^\text{su}_g$] start-up cost of the unit $g$ [COP]
\item[$c^\text{p}_g$] generation cost unit $g$ [COP/MW]
\item[$D_{t}$] demand during $t$ [MW]
\item[$RD_{g},RU_{g}$] ramp-down/up limit for $g$ [MW]
\item[$\overline{P}_g$] rated capacity for generator $g$ [MW]
\item[$P^\text{wind}_{t}$] wind generation during $t$ [MW].
\item
\item[\scalebox{1.2}{\textit{Variables}}]
\item
\item[$p_{g,t}$] generating power unit of $g$ during $t$ [MW]
\item[$\theta_{g,t}$] binary variable indicating if $g$ is on or off during $t$
\item[$\tau_{g,t}$] binary variable indicating if $g$ is being turned on during $t$.
\end{IEEEdescription}
\section{Introduction}\label{Sec: Intro}
At the end of the 20th century begins a shift in energy generation trends, a revolution promoted by concerns about pollution, climate change and the dependence of some countries on the import of hydrocarbons. In this context, considerable interest is provided to the opportunities offered by renewable energy technologies as a source of electric supply in non-interconnected areas and as part of the integrated generation system that participates in the electricity markets.

In general, policies of the different countries focus on promoting and increasing the proportion in the energy mix the following types of generation: solar photovoltaic, solar thermal, wind, geothermal, hydroelectric, wave, tidal and bio-fuel \cite{Hasani2012}. Different analyses of these technologies establish wind energy as having the highest returns on investment, and the most straightforward installation and maintenance procedures \cite{Baloch2016, Saidur2010}.

It is noteworthy that by the end of 2017 between only five countries had approximately 40 000 MW of newly installed power: China (19 500 MW), USA (7 017 MW), Germany (6 581 MW), Brazil (2 022 MW) and India (4 148 MW), which currently makes the wind energy the fastest growing generation technology at an annual growth rate of 23.6\%, with an installed capacity in the world of 540 GW by 2017 \cite{GWEC}.

In Colombia, the total installed capacity is 16.8 GW of which wind power generation currently represents only 0.1\% (19.5 MW) with a single wind park, Jep\'irachi \cite{UPME2015}. The Corporaci\'on Aut\'onoma Regional de La Guajira, through the Resolution No. 03357 of January 8 of 2010, granted the environmental license for the construction and operation of the 31,5 MW Jouktai wind park \cite{Corpoguajira}. The parks of Carrizal (195 MW), Casa El\'ectrica (180 MW) and Irraipa (99 MW) are expected to enter the same process, with a total of 505,5 MW of new wind power installed in the country. These projects are supported by the entrance of the Law 1715 whose primary objective is to promote the use of renewable sources in the Colombian electricity market and non-interconnected areas \cite{Law1715}. Similarly, the application of this law encourages the implementation of academic projects and research as in \cite{Restrepo2016}, where Restrepo et al. propose a method of evaluation and optimisation of self-generation from photovoltaic solar energy. Caicedo et al. analyse the impacts of the Law 1715 to stimulate the production of electricity by non-renewable sources in non-interconnected areas \cite{Caicedo2015}. Whereas, Manrique et al. present a technical and economic analysis of a photovoltaic solar energy system for residential users in the municipality of Ch\'ia (Cundinamarca, Colombia) \cite{RodriguezManrique2015}.

The main contribution of this work is the presentation of a comprehensive evaluation of the economic impact of incorporating wind generation in the Colombian electricity market. This is done through the simulation and analysis of the Colombian electric system in five steps: 
\begin{enumerate}[i]
    \item. \textit{Description of the Colombian electricity markets.} Based on the country's legislation the current market mechanisms and the regulations for the installation of non-conventional resources are outlined.  
    \item. \textit{Modelling of the Colombian spot market\footnote{The calculation of the \textit{spot price} is performed after the unit has produced the requested energy. Therefore the Colombian market is not a \textit{spot market} or generates a \textit{spot price}. The local terminology denominates both the market and calculated prices as \textit{spot} though. Therefore, the terms \textit{spot market} and \textit{spot price} will be used \cite{Camelo2018}.}.} The electricity market operation is presented in the basis of an optimisation model following the regulations established by the Energy and Gas Regulation Commission (CREG) and the restrictions of thermal generators. In this step, a unit commitment model of the generators that participate in the day-ahead dispatch is established.
    \item. \textit{Estimated wind generation.} In this stage, the wind power generated each hour is estimated from the historical wind data measurements for the future wind park.
    \item. \textit{Entry of the wind generation in the ideal pre-dispatch.} The unit commitment is solved for different values of installed wind capacity during selected typical days.
    \item. \textit{Assessment of the impact of wind generation.} The effects of the introduction of wind generation in the electricity market and power plant operation are analysed. 
\end{enumerate}

\section{The Colombian electric system: from the past to the present \label{Sec: System}}
\begin{figure}[h]
    \centering
    \includegraphics[width=0.9\columnwidth]{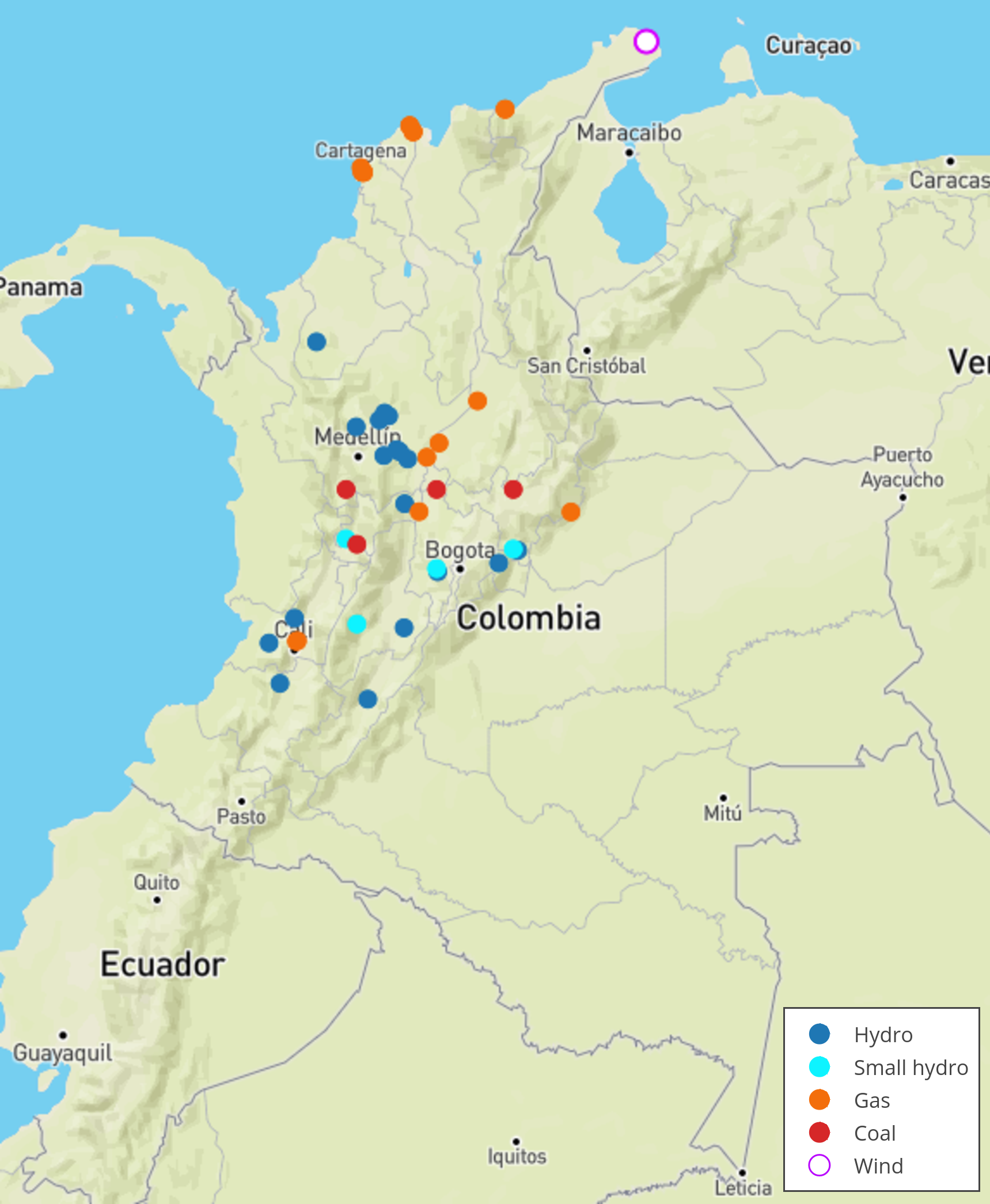}
    \caption{Location of the electric generation by technology in the Colombian territory. \label{Fig: Map}}
\end{figure}

Thanks to its location at the end of the Andes mountain range, Colombia has a significant number of rivers that descend from the mountain tops towards the Pacific and Atlantic ocean. The high density of rivers has allowed the country to capitalise the use of hydro power plants as its primary resource for electricity generation, concentrating most of these power plants at the foothills of the mountain range, i.e. the central region of the country (Fig. \ref{Fig: Map}). Large-scale hydro power (impoundment) provides 65\% of the installed capacity, whereas small hydro (diversion or run-of-river) accounts for 2.2\%. Gas and coal-fuelled generation is located in main urban centres and account, respectively,  for about 28 and 5\% of the total capacity. Other minor generators represent the remainder of the generation mix in smaller proportions.

\begin{figure}
    \centering
    \includegraphics[width=0.9\columnwidth]{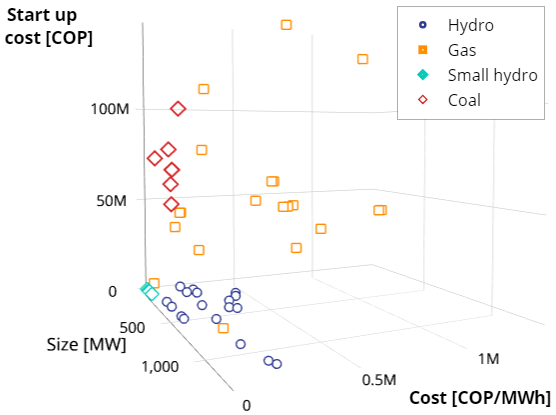}
\caption{Allocation of electric generation by technology in the Colombian power system. \label{Fig: Gen Description}}
\end{figure}

For their operation in the electricity market, large and small hydro power plants provide their start-up costs as zero (Fig. \ref{Fig: Gen Description}), varying their offered generation capacity based on the availability of water in the reservoir and the river flow, respectively. Small hydro plants bid the lowest marginal cost of generation since their power generation is continuous and dependent on the river flow. Fossil fuel plants have larger marginal and start-up costs, being employed for peak hours or as a backup generation in the presence of extended droughts that reduce the availability of the water resource, such as the \textit{ Ni\~no} weather phenomena \cite{Ealo2011}.
\subsection{Day-ahead market \label{Sec: Day-ahead}}
The electricity sector in Colombia had been characterised, up to the early 1990s, by independent state-owned regional and vertically integrated electricity companies which would exchange power through the national transmission system operated by the public company ISA (Interconexi\'on El\'ectrica S.A.) \cite{ISA}. After a period of economic instability and extended droughts brought by the \textit{Ni\~no} led to load severe shedding, the government issued the first laws towards the creation of a competitive electricity market. Through Law 142, known as the \textit{Utilities Law}, the existing monopolies for the electricity sector were dissolved. Whereas Law 143 of 1994, the \textit{Electricity Law}, sets the rules for the operation of the national interconnected system (SIN, in its Spanish acronym) under the regulation of the regulatory commission of energy and gas (CREG) based on a liberalised electricity market. The Colombian electricity market would start its operation in 1995.

Until 2001 the operation of the Colombian market was based on the self-scheduling of the power plants. The generation plants would send their scheduling for the next 24 hours based on three parameters for each one of these hours: i)  generated power, ii) price for the produced power, and iii) maximum generating capacity \cite{Camelo2018}. In 2001 it was required to only submit one price of generation for the whole 24 hours period.

Through the legal ruling 11 of 2009 and subsequent amendments, the CREG modified the operation of the electricity market. The modifications were aimed to improve the market efficiency through the inclusion of the start-up costs of the fuel-based power plants \cite{CREG011de2009}. The new electricity market would operate based on a centralised dispatch in which the generators would provide a 3-part hourly bid for the next 24 hours:
\begin{itemize}
    \item Start-up/shut-down costs submitted on a quarterly basis.
    \item Generation costs the same for the 24 hour period.
    \item Maximum available capacity different for each hour.
\end{itemize}
The centralised unit commitment serves as the current market mechanism, and it is controlled and performed by the market operator (XM, in its Spanish acronym). The dispatch of the units is based on their marginal cost of generation and through the compensation of their fixed costs. Hence, the market spot price is calculated as a function of the clearing price (cost of the marginal dispatched unit) and the start-up costs of the generators (to be compensated).

Generators are divided into three groups based on their rated capacity and participation in the spot market. Plants above 20 MW must participate in the electric spot market. Whereas, plants between 10 and 20 MW can choose whether to participate in the market. Finally, those below 10 MW act as price takers based on their available capacity.
\subsection{Capacity market}
The same weather phenomena that motivated the creation of the spot electricity market prompted the creating of a reserve market to increase the reliability of the system. The occurrence of the \textit{Ni\~no} would decrease the water availability for hydro-power generation, causing a crisis in the electricity sector due to the significant share of the electric generation mix corresponding to this resource. Nation-wide electricity rationing, through power curtailments, was adopted in the country, alongside to shifting of the time zone for one hour. Such measures lasted from March 1992 until February 1993.

In order to prevent future electricity rationing, given the cyclic nature of the \textit{Ni\~no}, the Law 143 of 1994 article 23 introduces the \textit{capacity charges}. The capacity charge was a payment made to fossil-fuel power plants based on their installed capacity. These plants were to be used when the hydro reserves were below an established limit. With the introduction of the CREG legal ruling 71 of 2006, the capacity charges were redefined as \textit{reliability charges} with the auctioning of Firm Energy Obligations (OEF). The OEFs are reserve capacities offered by the generators in the system and auctioned regarding their capacity and duration. The generators that receive OEFs receive a fixed payment, relative to the committed capacity, for the validity period of the OEF. This payment is received even if the reserved capacity is not employed.

The activation mechanism for the OEFs is based on the behaviour of the electricity spot price relative to the scarcity price. The scarcity price is defined by the CREG on a monthly basis as a function of the current and projected fuel prices. If the market spot price surpasses the scarcity price, at least 1 hour of the day, the plants with OEFs must generate electricity based on their committed obligations. Additional to the fixed amount received when providing the OEF, the generation plants collect an additional payment depending on the amount of energy provided during the activation of the OEF. If the provided energy is equal to the established in its OEF the plant is paid at the scarcity price. Whereas if the generated energy exceeds the OEF, the surplus of energy is paid at the spot price. In this way, the spot price would not increase too much during drought periods, and the installation of more efficient fossil-fuel generation is supported.

The power plants with OEFs must guarantee their obligation fulfilment through the assurance of their fuel contracts. In case of not being able to comply with the OEF, the plant can acquire the remaining energy through the established security rings\footnote{ The secondary markets are based, in their different mechanisms, in bilateral contracts between market agents (generators and retailers) and are used for the realisation of the spot market and OEF commitments.}. If the power plant fails to do so, it must compensate the curtailed demand.
\subsection{International connections}
The international interconnections of the country are at 230kV to the east with Venezuela via bilateral markets (capacity for export - 285MW and import - 215MW), and to the south-west to Ecuador through implicit auctions (export - 336MW and import - 205 MW) \cite{XM}.

Additionally, an interconnection project to sell electricity surplus from Colombia to Panama was started in 2007 but was later abandoned due to the introduction of liquefied fuels in the Panamanian generation matrix. The project consisted of a 400 MW HVDC line of 500 Km, allowing the interconnection with the capital of the neighbouring country and also setting a first step towards the interconnection with Central American countries \cite{Panama}. The national governments are currently negotiating the reactivation of this project.
\subsection{Introduction of non-conventional sources}
In 2014 with the objective to diversify the generation matrix, the Colombian government introduced the Law 1715 of 2014, known as the "Renewable Energy Law" (REL) \cite{Law1715}. The REL promotes the use of the following generation and efficiency mechanisms:
\begin{itemize}
    \item Non-Conventional Renewable Energy (FNCER): biomass, small hydro (run-of-river), wind, geothermal, solar and tidal.
    \item Efficient energy management.
    \item Demand response.
    \item Self-generation.
    \item Distributed generation.
    \item Replacement of diesel generation.
\end{itemize}

The REL establishes responsibilities for the Program for the Rational and Efficient Use of the conventional and non-conventional forms of energy (PROURE) designed by the Ministry of Mines and Energy in 2001 (Law 697 of 2001) \cite{Law697}. The PROURE must set the goals and incentives for the replacement in public utilities of outdated equipment by modern ones with higher energy efficiency. Additionally, the Program will establish step-wise goals spanning ten years for the improvement of the energy efficiency in public buildings.

The REL places the responsibility on the CREG for devising the mechanisms for the implementation of Demand Response (DR). Such mechanisms were outlined in the Statute 011 of 2015 \cite{011of2015}. The users willing to participate in DR must send through their retailer a two-part bids to the manager of the system for commercial exchanges (ASIC) consisting of the fixed price of reduction for the next 24 hours (unique value in integer values of COP/MWh) and the maximum reduction commitment (in an hourly basis and with integer values of MW). If the user is required to change its consumption, based on the offered DR, then it will be paid the agreed price plus the difference between the scarcity price and the spot price for the relevant hours.   

For small distributed generators that do not participate in the spot market, their surplus of energy can be delivered to their distribution or transmission network. With the use of a bidirectional meter, energy credits will be awarded to the power plant, and those could be later sold or negotiated. On the other hand, distributed generators will be paid based on the benefits provided to their distribution networks, such as reactive power support, improvement of the life-cycle of their distribution equipment, and reduction of losses. The substitution of diesel generators for FNCER in areas not served by the SIN, will be awarded economic incentives in correspondence to the obtained improvement in efficiency. The main benefits for the installation of FNCER can be summarised as:
\begin{enumerate}[i]
    \item. Reduction for five years of income and property taxes by 50\% to the investors.
    \item. No VAT on acquired services or equipment investments.
    \item. No import tariffs for equipment investments.
    \item. Accelerated depreciation of the equipment.
\end{enumerate}

\section{Methodology}
\subsection{Colombian ideal pre-dispatch} \label{Sec: Pre-dispatch}
The operation of the National Interconnected System (SIN) and the administration of the Colombian electricity market are carried out by the XM \cite{XM}, per the regulations established by the CREG. XM receives the offers of variable costs of the generators for the matching of supply and demand using the single price model and merit order for the generation schedule. In Colombia, generation and commercialisation of electric power are governed by a framework of competition in the liberalised market, while transmission and distribution are natural monopolies regulated by the CREG. The commercialisation of energy is carried out through a short-term market (energy stock exchange) and a long-term market (contracts). The implemented unit commitment model, Model \ref{Mod: UC Col}, is based on the market rules outlined in Section \ref{Sec: Day-ahead}. The impact of wind generation is analysed for installed capacities between 0 MW (current scenario) and 1 000 MW (high wind installation).

The objective of the unit commitment is to minimise the operational cost of the generators while satisfying the demand during a given day. Expression \eqref{Eq: Objective M1} provides the total operational cost of generation, where the first term represents the costs associated with the power generation and the second one the costs related to the start-up of the power plants. The power balance is expressed in \eqref{Eq: P Balance}, where the demand at every time step, $D_t$, must be equal to the summation of the power generated by the wind turbines, $P^\text{wind}_t$, and that produced by the generation units, $p_{g,t}$. For this model the demand is assumed constant, given their low elasticity in the Colombian market (between -0.067 and -0.12) \cite{Gutierrez2011}. Additionally, the wind is a non-dispatchable resource (cost zero). The capacity limits, \eqref{Eq: P Limits}, provide the technical limit for the maximum generation from a power plant $g$. While \eqref{Eq: Ramps} models the maximum change in generated power between consecutive hours. Inequalities \eqref{Eq: Start1}-\eqref{Eq: Start3} allow to represent the start up process of unit $g$.
\begin{model}[h]
\caption{Unit commitment model}
\vspace{-0.3cm} 
\label{Mod: UC Col}
\begin{subequations}  \label{eq: PWModel}
\begin{IEEEeqnarray}{ll}
\noalign{\noindent \bf Indices: \vspace{\jot}}
g \;\;   &\mbox{ - generation unit}  \nonumber\\
t \;\;   &\mbox{ - time step \quad [h]}  \nonumber\\
\noalign{\noindent \bf Parameters: \vspace{\jot}}
c^\text{su}_g \;\;   &\mbox{ - start up cost unit $g$ \quad [COP]}  \nonumber\\
c^\text{p}_g \;\;   &\mbox{ - generation cost unit $g$ \quad [COP/MW]}  \nonumber\\
D_{t} \;\;   &\mbox{ - demand during $t$ \quad [MW]}  \nonumber\\
RD_{g} \;\;   &\mbox{ - ramp-down limit for $g$ \quad [MW]}  \nonumber \\
RU_{g} \;\;   &\mbox{ - ramp-up limit for $g$ \quad [MW]}  \nonumber \\
\overline{P}_g \;\;   &\mbox{ - rated capacity for generator $g$ \quad [MW]}  \nonumber \\
P^\text{wind}_{t} \;\;   &\mbox{ - wind generation during $t$ \quad [MW]}  \nonumber\\
\noalign{\noindent \bf Variables: \vspace{\jot}}
p_{g,t} \;\;   &\mbox{ - generating power unit $g$ during $t$ \quad [MW]}  \nonumber \\
\theta_{g,t} \;\;   &\mbox{ - binary variable indicating if $g$ is on or off during $t$}  \nonumber \\
\tau_{g,t} \;\;   &\mbox{ - binary variable indicating if $g$ is being turned on}\nonumber \\
&\mbox{during $t$}  \nonumber
\end{IEEEeqnarray}
{\bf Objective}:
\begin{IEEEeqnarray}{ll}
\qquad \minimize \; \sum\limits_{g,t}\ \big[ c^\text{p}_g \cdot p_{g,t} + c^\text{su}_g \cdot \tau_{g,t} \big], &\forall g,t
    \IEEEyessubnumber* \IEEEeqnarraynumspace \label{Eq: Objective M1}\\
\noalign{\noindent \bf Constraints: \vspace{2\jot}}
D_t = {\sum\limits_{g}}p_{g,t} + P^\text{wind}_{t}, &\forall t \IEEEeqnarraynumspace \label{Eq: P Balance}\\
\theta_{g,t} \cdot \underline{P}_g \le p_{gt} \le \theta_{g,t} \cdot \overline{P}_g, & \forall g,t \IEEEeqnarraynumspace \label{Eq: P Limits}\\
-RD_g \le p_{g,t}-p_{g,t-1} \le RU_g, \qquad \qquad\qquad & \forall g,t \IEEEeqnarraynumspace \label{Eq: Ramps}\\
\tau_{g,t} \ge \theta_{g,t} - \theta_{g,t-1}, & \forall g,t \label{Eq: Start1}\IEEEeqnarraynumspace\\
\tau_{g,t} \le 1 - \theta_{g,t-1}, & \forall g,t\IEEEeqnarraynumspace\label{Eq: Start2}\\
\tau_{g,t} \le \theta_{g,t}, & \forall g,t. \IEEEeqnarraynumspace\label{Eq: Start3}
\end{IEEEeqnarray}
\end{subequations}
\end{model}
\subsection{Characterisation of the wind generation}
The calculation of the electric power production of a wind farm has been carried out in three stages. As a first step, it is necessary to record wind speeds at the selected site during a sufficiently characteristic time, for the analysis and estimation of the wind resource. Secondly, the appropriate type of turbine for the available resource is selected. Lastly, with the characteristics of the chosen turbine and the expected wind speeds, the production of electric energy is calculated.
\subsubsection{Characterisation of the wind resource}
The characterisation of the wind resource can be done through meteorological variables or statistical variables based on the recorded measurements. The study of meteorological variables considers the correlation of the wind speed with other climatic factors such as temperature and atmospheric pressure, which leads to a formulation with an extensive set of variables.

\begin{figure}
\centering
\includegraphics[width=0.9\columnwidth]{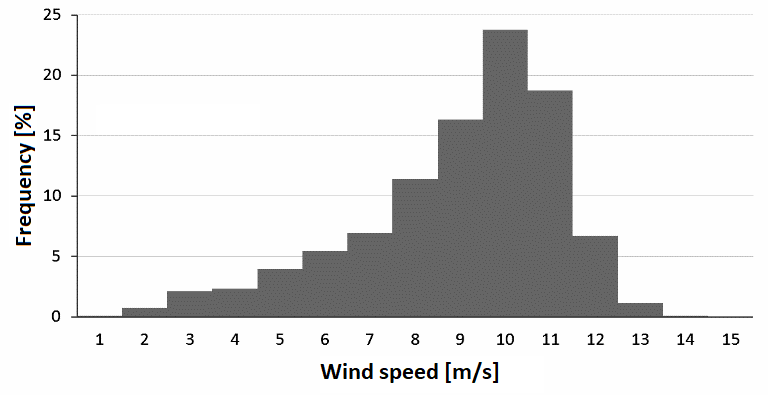}
\caption{Wind speed histogram \label{Fig: Histogram}}
\end{figure}

On the other hand, the study of statistical variables considers the speeds recorded as a time series and allows to model the wind behaviour based on statistical tools (frequency tables and histograms). For the present study, wind speed measurements obtained at a site located at 71.99 degrees west and 12 degrees north during 24 hours have been used for a year of study. An examination of descriptive statistics has been carried out, and the measures of central tendency and dispersion for the annual data set have been calculated. Table \ref{Tab: Wind statistics} summarises the data obtained for the considered year.
\begin{table}
\caption{Wind speed data summary\label{Tab: Wind statistics}}
{\begin{tabular*}{20pc}{@{\extracolsep{\fill}}lc}
\toprule
 & \textbf{Wind speed [m/s]} \\
\midrule
Mean & 8.49 \\
Median & 9.03 \\
Standard deviation & 2.24 \\
Minimum & 0.25 \\
Maximum & 13.45 \\
\hline
\end{tabular*}}{}
\end{table}
Each of the measurements in Table \ref{Tab: Wind statistics} provides information about the wind at the site studied. A preliminary model can be constructed from the obtained mean speed and its standard deviation. The mean represents the characteristic value of wind speed, while the standard deviation gives a measure of the variability of the data. In the histogram of Fig. \ref{Fig: Histogram} the frequency of occurrence of wind speed records can be observed.

\subsubsection{Selection of the type of turbine}
Once the wind resource on the site has been characterised, the type of turbine is classified using the IEC 61400-1 standard \cite{IEC2005}. The site studied corresponds to a class II turbine with an average annual speed of 8.49 m/s. The selection criteria for the wind turbine model were two: the safety of the turbine, considering the maximum wind speed, and the maximisation of the energy produced. The turbine selected is the V90-1.8 / 2MW from the Danish company Vestas \cite{Vestas2018}, which has an IIA classification according to the IEC standard, a cut-in and cut-out wind speeds of 4 and 25m/s, respectively (Fig. \ref{Fig: Turbine curve}).
\subsubsection{Energy production of the wind farm}
\begin{figure}
\centering
\includegraphics[width=\columnwidth]{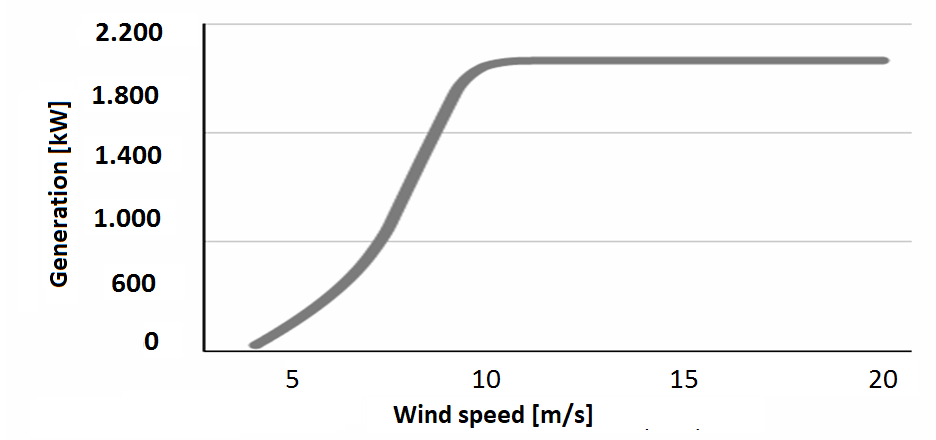}
\caption{Power curve for the wind turbine Vestas V90/2MW \cite{Vestas2018}. \label{Fig: Turbine curve}}
\end{figure}

Once the operational characteristics of the selected turbine are known, and with the information of the wind resource, the production of electrical energy is estimated. For the evaluation of the generated electrical energy, static analysis is used, which ignores the periods of maintenance or other phenomena that may present a decrease in production. The static analysis is carried out from the hourly wind speed records (Fig. \ref{Fig: Histogram}) and the power curve of the wind turbine (Fig. \ref{Fig: Turbine curve}).
\section{Results}
The analysis of the Colombian electricity market with wind penetration is performed for 25 typical days with hourly demand and for different values of wind installation, ranging from 0 MW to 1 000 MW. In order to ease the evaluation of the different scenarios of installed wind capacity, three cases will be highlighted: \CaseA -- 0 MW, \CaseB -- 505.5 MW, and \CaseC -- 1 000 MW. \CaseA, and \CaseB{} respectively correspond to the current system operation and the operation with the approved projects. \CaseC{} represents an expected future installation value.

\begin{figure}
\centering
\includegraphics[width=0.9\columnwidth]{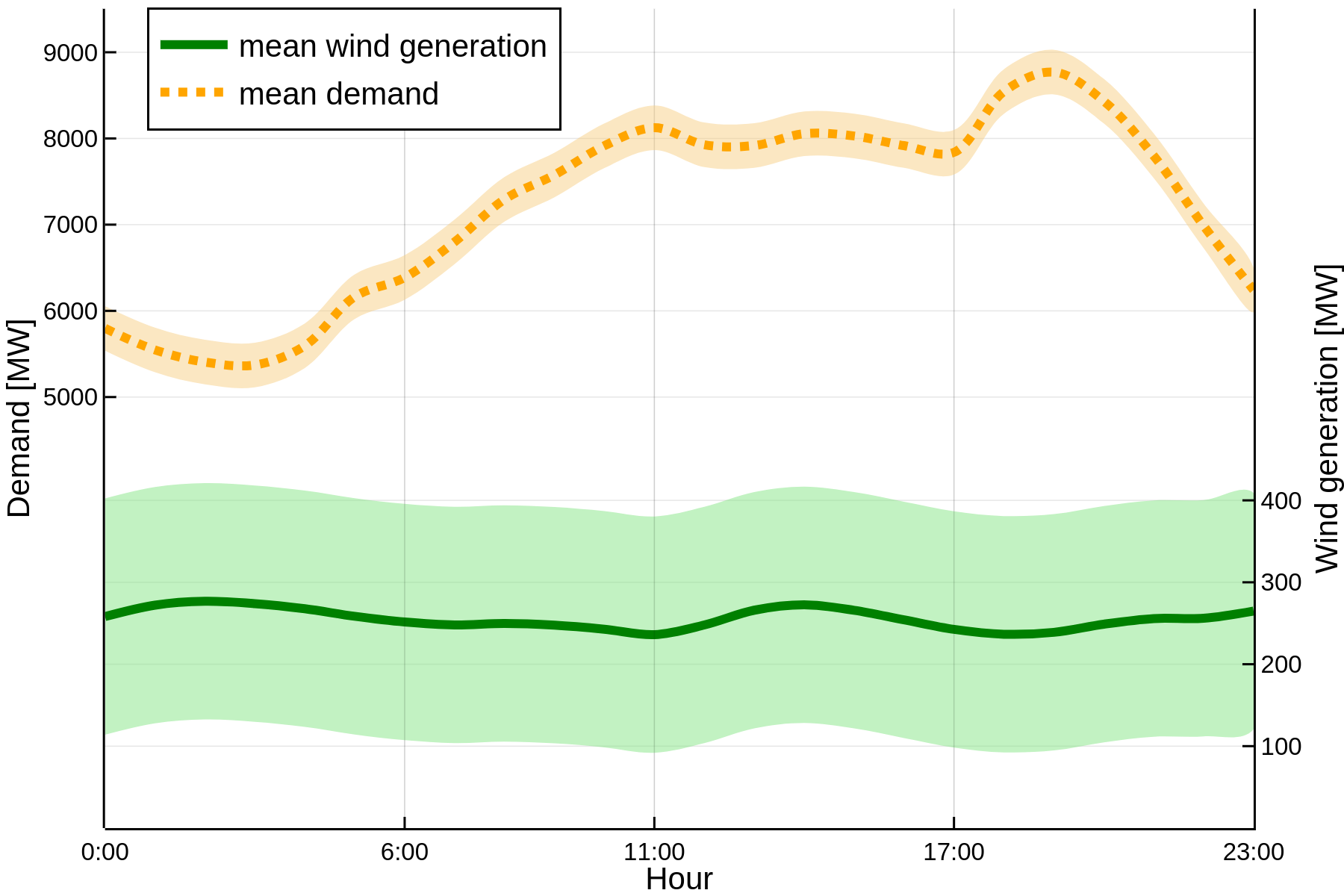}
\caption{Average annual demand and wind generation for \CaseB. \label{Fig: Demand_Wind}}
\end{figure}

The average hourly demand and wind generation, for \CaseB{} (505.5 MW), for the typical days is presented in Fig. \ref{Fig: Demand_Wind}. No complementarity of the wind generation and the demand is present since the average wind power remains between 230 and 280 MW without pronounced peaks. The demand and wind variability, represented by their standard deviation as the shaded area, remains relatively constant throughout the day. The hourly demand and wind generation for \CaseB{} for the typical days, as well as the technical data for the generation units, are given in the online dataset \cite{GonzalezWindData}.

The market spot price, paid by the demand, is calculated for each hour based on the price of the marginally dispatched unit ($\mathit{MPO}_t$), and a price uplift ($P^\text{up}$), introduced to compensate the start-up costs \cite{Camelo2018}. Expression \eqref{Eq: SpotPrice} allows to calculated the market spot price: 
\begin{IEEEeqnarray}{rCll}
P_t &=& \mathit{MPO}_t + P^\text{up}, &\forall t \label{Eq: SpotPrice}\\
\noalign{\noindent where 
\vspace{2\jot}}
P^\text{up} &=& \frac{\sum_g \max\{0,C_g^\text{MPO} - C^\text{plant}_g\}}{\sum_t D_t} \qquad &\IEEEyessubnumber*\\
\noalign{\noindent and
\vspace{2\jot}}
C_g^\text{MPO} &=& \sum_t \mathit{MPO}_t \cdot p_{g,t}, &\forall g\\
C_g^\text{plant} &=& \sum_t c_g^\text{p} \cdot p_{g,t} + c_g^\text{su} \cdot \tau_{g,t}, &\forall g.
\end{IEEEeqnarray}

All of the generation agents are paid the spot price $P_t$ for every unit of energy produced. The units whose $C_g^\text{plant} \ge C^\text{MPO}_g$ must reimburse the uplift payment ($P^\text{up}\cdot p_{g,t}$), while the other plants (whose start-up costs is not be covered by the $\mathit{MPO}_t$) must not reimburse any of the received payment. In this manner, the generation units with start-up costs are compensated.

The unit commitment described in \ModelOne, as well as the calculation of the spot price \eqref{Eq: SpotPrice} is performed independently for each of the selected days and wind penetration values. The presented simulations are performed using the modelling software Julia 0.6.3 \cite{Bezanson2017}, JuMP 0.18 \cite{Lubin2015}, with Gurobi 7.5.1. \cite{gurobi} as a solver.
\subsection{Impact of wind generation in the spot prices}
\begin{figure}
    \centering
    \includegraphics[width=0.9\columnwidth]{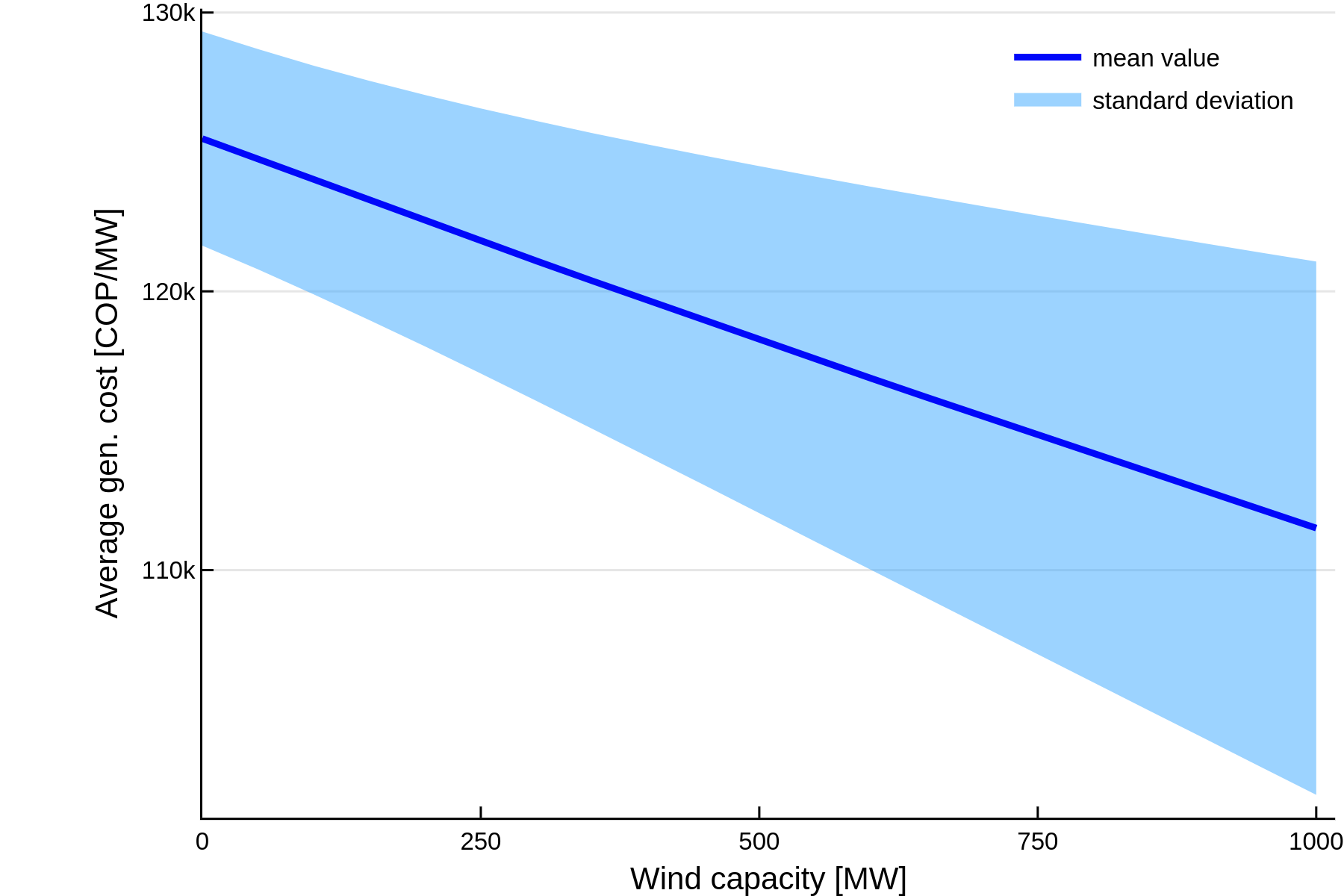}
\caption{Average daily generation cost vs. installed wind capacity. \label{Fig: Average Cost vs Wind}}
\end{figure}

The introduction of the wind power in the economic dispatch at zero marginal cost reduces the electricity generation costs. The decrease in the spot prices is proportional to the installed wind capacity. Fig. \ref{Fig: Average Cost vs Wind} provides the change in the daily average generation cost for different installed wind capacities. The solid line gives the mean value between the analysed scenarios, while the filled area represents its standard deviation. As seen in the figure, the price variability increases for higher values of installed wind capacity; corresponding to a more significant fluctuation of the generated wind power.

The average ($c^p_\text{mean}$), minimum ($c^p_{\min }$) and maximum ($c^p_{\max }$) average generation cost of electricity were calculated for the current, planned and high installed capacity; 0, 505.5 and 1 000 MW, respectively (Table \ref{Tab: Average Cost Min_Mean_Max}).  Even though the average and minimum generation cost decrease proportionally to the installed wind capacity, the maximum decreases in a smaller proportion; as a result of the increase in the cost variability for larger installed capacities of wind (Fig. \ref{Fig: Average Cost vs Wind}).

\begin{table}
\caption{Average electricity cost by installed capacity\label{Tab: Average Cost Min_Mean_Max}}
{\begin{tabular*}{20pc}{@{\extracolsep{\fill}}lccc}
\toprule
Installed wind capacity & $c^p_{\min}$ & $c^p_\text{mean}$ & $c^p_{\max }$ \\
                        & [COP/MWh] & [COP/MWh] & [COP/MWh] \\
\midrule
Case A (0 MW)         & 111 849.0 &  125 477.5 & 129 949.0 \\
Case B (505.5 MW)     & 101 497.7 &  118 265.5 & 129 152.5 \\
Case C (1 000 MW)     &  91 304.9 &  111 504.9 & 128 469.3 \\
\hline
\end{tabular*}}{}
\end{table}

\begin{figure}[!h]
\centering
\includegraphics[width=\columnwidth]{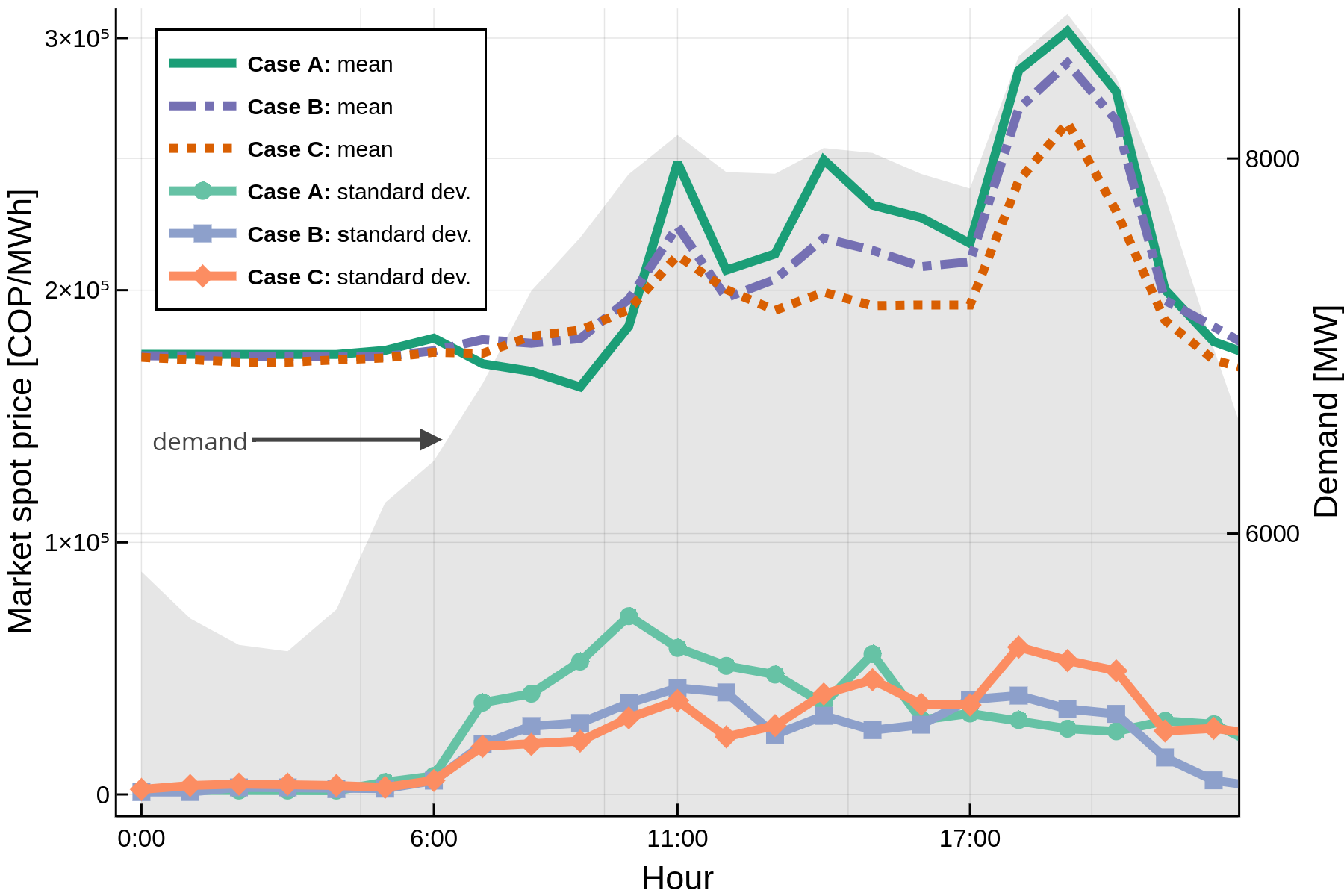}
\caption{Market spot price: \CaseA{} (0 MW), \CaseB{} (505.5 MW), and \CaseC{} (1 000 MW). \label{Fig: Spot Price}}
\end{figure}

The average spot price for 24-hours of the selected typical days was calculated at different values of wind capacity (Fig. \ref{Fig: Spot Price}). During the first six hours of the day, the spot price remains relatively constant for the different scenarios of installed wind. The invariability of the market price during these hours corresponds to the use of base generation.

For hours between 06:00 and 10:00 the average price for \CaseA{} diminishes and is lower than that of \CaseB{} and \CaseC{}. This decrease in price value for \CaseA{} corresponds to the disconnection of a gas generator at the sixth hour, to allow hydro power plants to cover the steeper demand ramp starting at 06:00. Whereas for \CaseB{} and \CaseC{}, the wind generation allows compensating the increased demand change. The variability during this period appears to be reduced with the introduction of wind generation, evidenced by a lower standard deviation for \CaseB{} and \CaseC{}.

During the remaining hours, there is a higher reduction of the spot price for \CaseB{} and \CaseC{}, when compared to \CaseA{}. The price reduction is especially more significant during the high demands, given the decreased use of the peaking units. The variability of the wind power generation increases during these hours, resulting in an increase of reserve requirements \cite{Dvorkin2015}.

\subsection{Effect on generation share and generator's revenue} \label{Sec: Change Share}
\begin{figure}
    \centering
    \includegraphics[width=0.9\columnwidth]{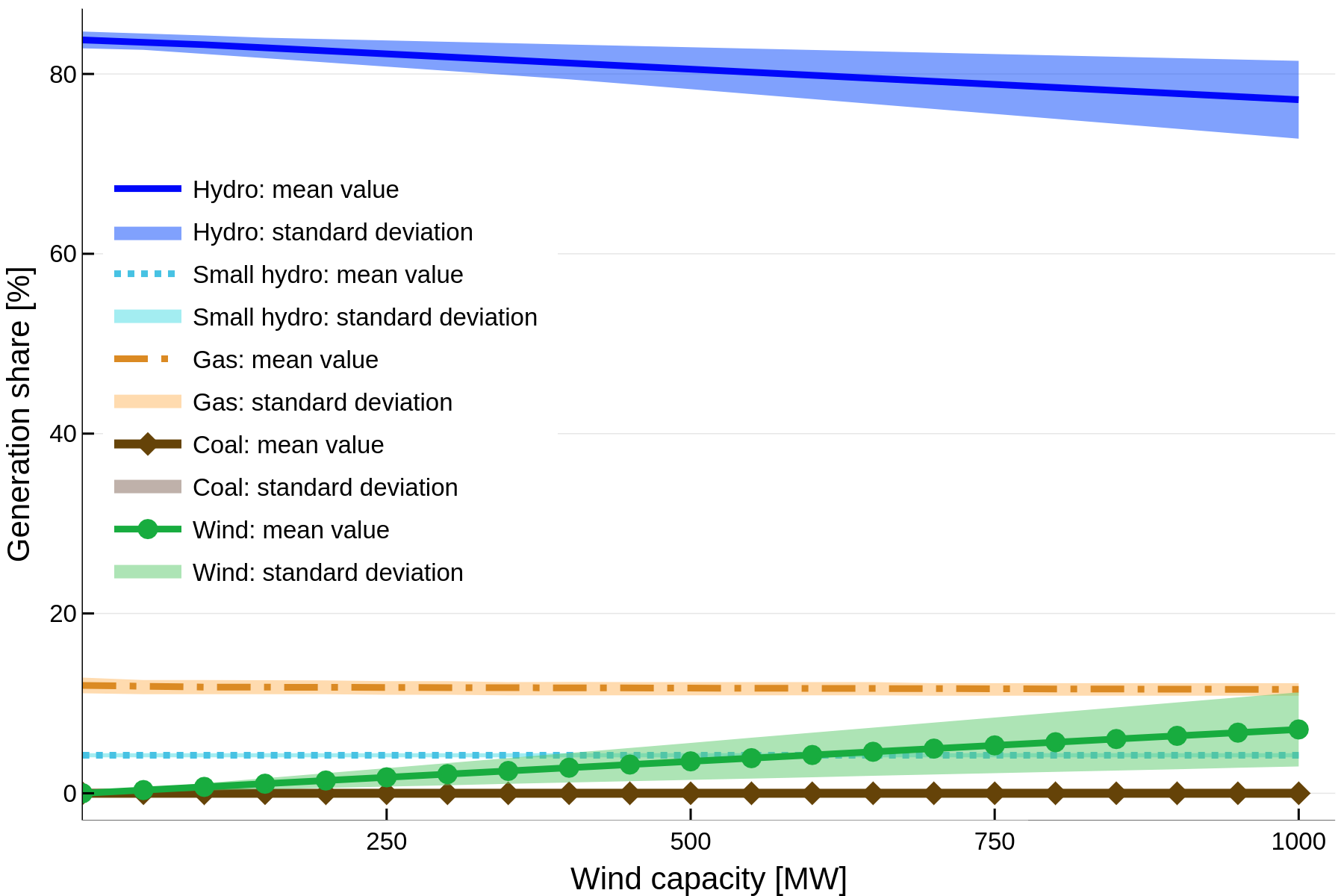}
\caption{Generation share by type.
\label{Fig: Share by type}}
\end{figure}

The generation share by type of power plant for different wind capacities can be seen in Fig. \ref{Fig: Share by type}. The solid line represents the average share in the selected scenarios and the filled area the standard deviation. As seen in the figure, the share of wind generation increases with its installed capacity as expected, but also does its variability. The coal plants are not dispatched in the unit commitment, while the small hydro plants keep their share constant with the increase of wind penetration, i.e. they are always dispatched. The gas-fired plants have a slight decrease in their share of the mix with a small deviation. The most significant change in the generation matrix occurs for the hydro power plants, with a continued decrease as more wind is injected into the system. The hydro generation share also increases its variability with the increased wind. The greater variation of hydro production with higher wind penetration means that the hydro plants are being used to compensate for the fluctuations in the wind generation. This result is counter-intuitive since the hydro power plants posses a smaller generation cost than the ones based on fossil fuels. The use of the hydro plants in this manner can be explained by their zero start-up costs, leading to the higher switching of these units. It is more economically efficient to turn on the gas generators only once per day to cover the peak requests and switch on and off the hydro plants for smaller fluctuation in demand and wind.

\begin{figure}
    \centering
    \includegraphics[width=0.9\columnwidth]{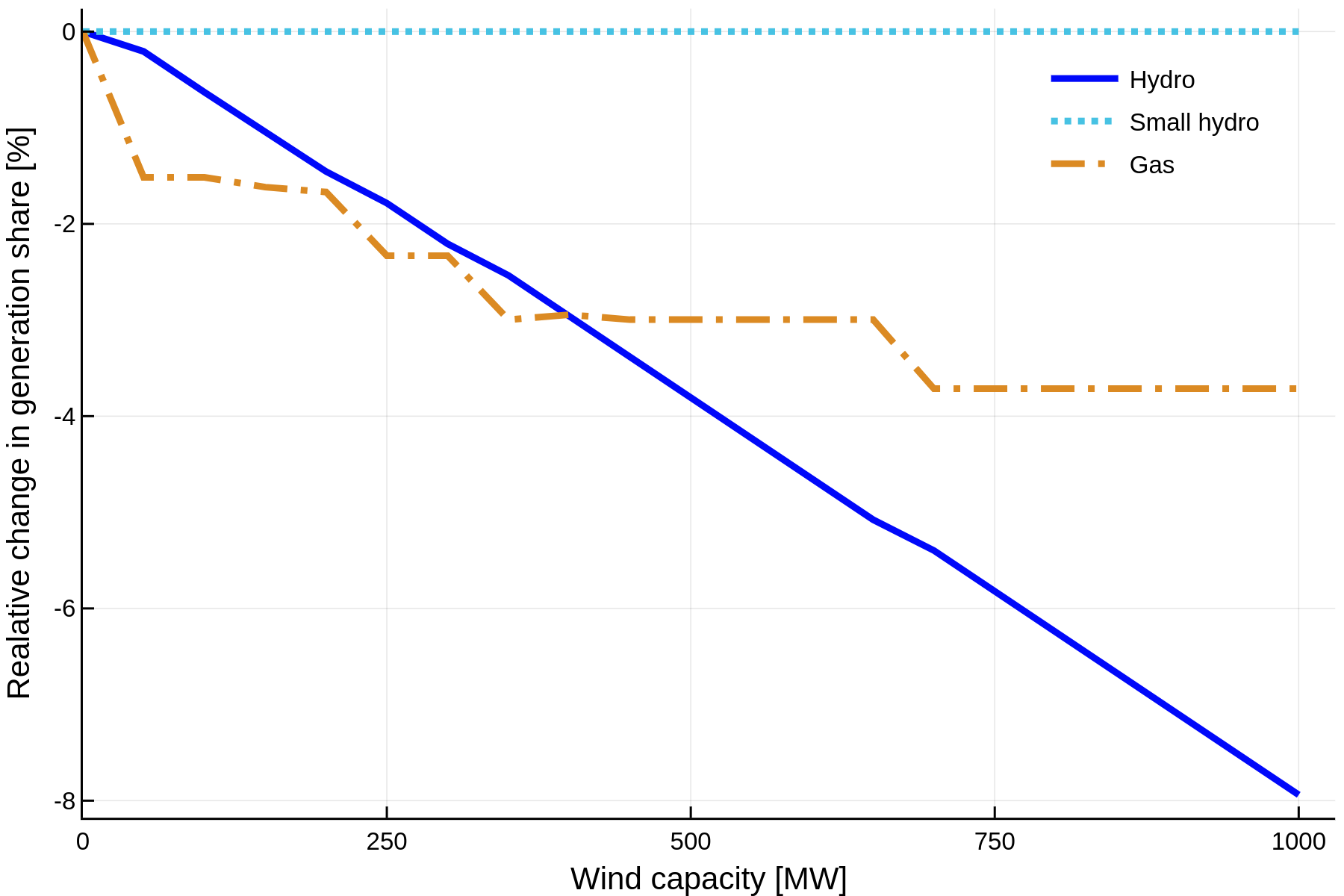}
\caption{Generation share change vs. installed wind capacity. \label{Fig: Change Share}}
\end{figure}

The average change in the generation share by plant type, in comparison to the current state of no wind generation, is presented in Fig. \ref{Fig: Change Share}.  The decrease in the share of hydro generation is proportional to the installed wind capacity. Whereas, the amount of gas-fuelled power decreases in steps. This step-wise decrease corresponds to the ability of the wind generation to cover a higher percentage of the demand. 

\begin{figure}
    \centering
    \includegraphics[width=0.9\columnwidth]{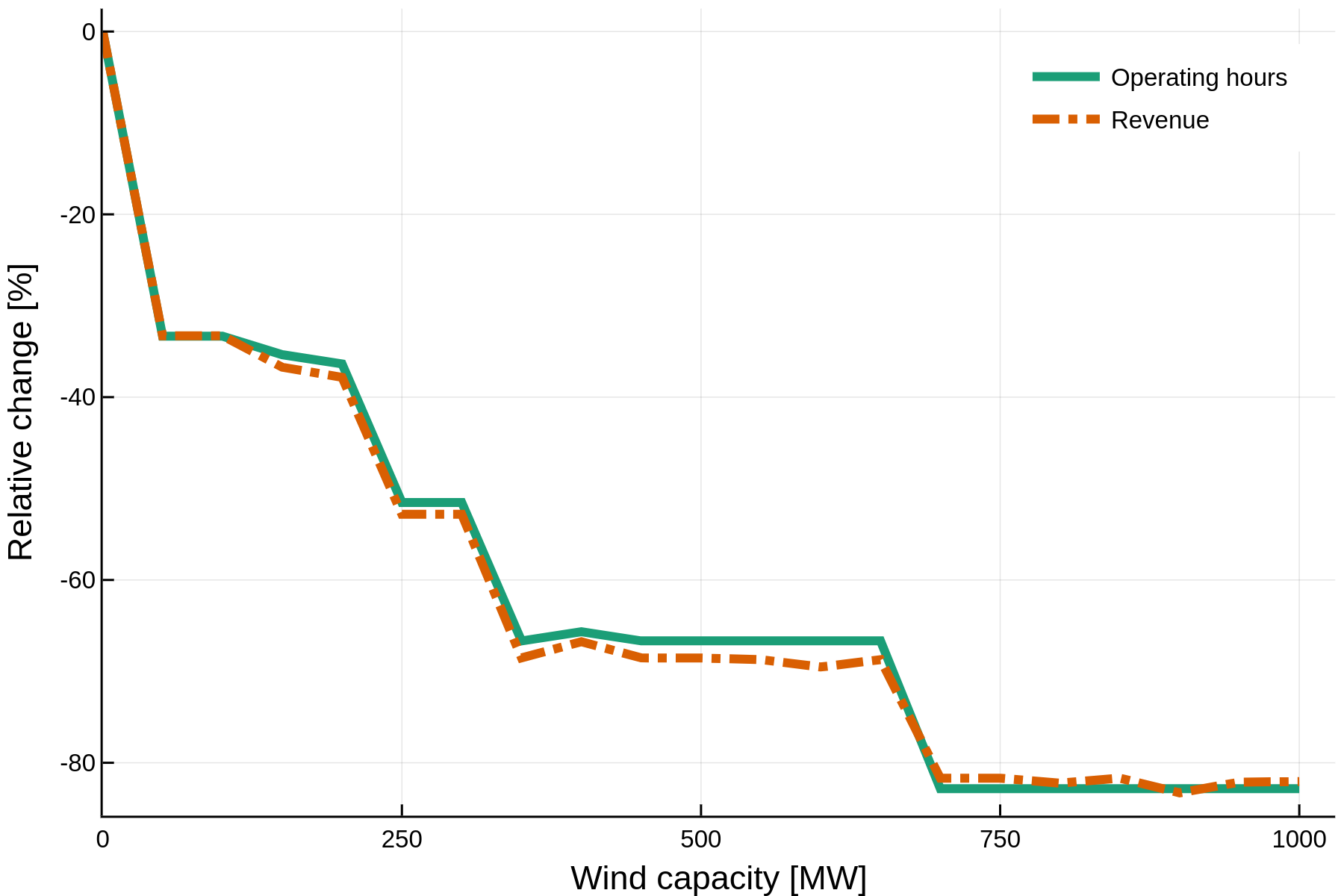}
\caption{Change in operating hours and revenue for generator 40. \label{Fig: Change Revenue}}
\end{figure}

When analysing the gas plants operating during the unit commitment, it can be seen that only 3 out of 20 are dispatched. Two of these plants, generator 28 and 47, work as base generation during all of the selected scenarios. Meanwhile, generator 40 decreases its number of operating hours with the increase of wind generation (Fig. \ref{Fig: Change Revenue}). The plant's revenue is proportional to the number of operating hours, dropping as low as 20\% of its original value for an installed wind capacity of 1 000 MW. Given that gas power plants usually operate with a profit margin of 11\% \cite{Simshauser2014}, such a drastic decrease in operating revenue could lead to a permanent loss of economic profitability for the plant. Thus, forcing the plant to operate through more volatile streams of revenue such as reliability charges and secondary markets, if decommissioning is to be avoided.

\section{Discussion, recommendation and conclusion}

In this paper, we discussed some of the potential effects of new investment on renewable generation, wind in particular, in the Colombian power system under the new regulatory policies recently approved.  For doing it, we have presented in a simplified way the regulatory evolution to understand the current state of the Colombian power system. Like many other Latin American countries, Colombia has an essential share of hydro-power generation which is strongly dependent on meteorological cycles as the Ni\~no.  The recently approved Law 1715 would facilitate the increase of the renewable generation share in the coming years. The current market design for the day-ahead power wholesale is based in a cost-audited unit commitment with no consideration of network constraints.  Under such a framework, we have simulated and analysed the economic impact of different wind generation in the Colombian electricity market. 

We found that an increase in the wind resource penetration would lead to a dire diminish in the number of operating hours, and proportionally its revenue, for some gas-fired power plants. Such a change in the operation regime of the plant would lead to an unfeasibility of its economical operation and eventual decommissioning.  Additionally, it is expected to have a general reduction of the spot price while its variability would potentially be increased. Thus, increasing the need for reserve allocation. However, wind and hydro technologies have shown complementary in the operation of the system.  Thus, the hydro-generator could serve to alleviate wind generation variability.

Further research is needed in the context of the Colombian power system. Several aspects should gain attention as long as the renewable generation increases.  
\begin{itemize}
    \item Firstly, it is important to revise the unit commitment structural model that currently is used for the day-ahead power exchange.  Some of the limitations of the current model are i)  Transmission constraints are not considered. It is important to understand that complementarity for balancing wind intermittence by using reserves from hydrogeneration requires ``deliverability'' guarantees; ii) Reserve allocation is not considered in the day-ahead dispatch. Energy and reserve co-optimization have demonstrated advantages in the North-America power systems concerning the energy-only markets; iii) Pricing mechanism should be revisited accordingly; iv) Also, bid-based instead cost-based offers should be analysed in the new context. 
    \item Secondly, the impact of climate vulnerabilities such as climate cycles represents another challenge for Colombia with a significant share of hydro generation. It has been observed that there exists a correlation between the wind speed in the North of the country and the presence of the Ni\~no \cite{Ealo2011}. This fact could be seen as an opportunity.
    \item Finally, we recommend policymakers and regulators to give greater attention to the economic impact of green policies on existing fuel-based power plants. Especially considering the reliability services provided by these units in years of hydro-generation resource scarcity. Thus, long-term climate-aware studies are needed in this context of security and system reliability.
\end{itemize}

\bibliographystyle{IEEEtran}
\bibliography{WindRef}

\end{document}